\documentclass[reprint,amsmath,amssymb,aps,prl,]{revtex4-1}

\usepackage{amssymb}
\usepackage{amsmath}
\usepackage{gensymb}
\usepackage{bm}
\usepackage{bbm}
\usepackage{graphicx}
\usepackage{dcolumn}
\newcommand{\be}{\begin{eqnarray}}
\newcommand{\ee}{\end{eqnarray}}
\newcommand{\la}{\langle}
\newcommand{\ra}{\rangle}
\newcommand{\tr}{{\rm Tr}}
\newcommand{\one}{\mathbbm 1}
\newcommand{\tht}{\tfrac{\theta}2}

\begin{document}

\title[On the Origin of Quantum Uncertainty]{On the Origin of Quantum Uncertainty}
\author{Christoph Adami}
\affiliation{Department of Physics and Astronomy, Michigan State University, East Lansing, MI 48824}

\begin{abstract}
The origin of the uncertainty inherent in quantum measurements has been discussed since quantum theory's inception, but to date the source of the indeterminacy of measurements performed at an angle with respect to a quantum state's preparation is unknown. Here
I propose that quantum uncertainty is a manifestation of the indeterminism inherent in mathematical logic. By explicitly constructing pairs of classical Turing machines that write into each others' program space, I show that the joint state of such a pair is determined, while the state of the individual machine is not, precisely as in quantum measurement. In particular, the eigenstates of the individual machines appear to be superpositions of classical states, albeit with vanishing eigenvalue. Because these ``classically entangled" Turing machines essentially implement undecidable ``halting problems", this construction suggests that the inevitable randomness that results when interrogating such machines about their state is precisely the randomness inherent in the bits of Chaitin's halting probability. Because this classical construction mirrors quantum measurement, I argue that quantum uncertainty has the same origin.
\end{abstract}

\keywords{quantum measurement, halting problem, quantum uncertainty, classical indeterminacy}



\maketitle


Quantum physics is weird. When we measure a quantum state, the result is uncertain unless we previously prepared the state exactly the way we were going to measure it. When we measure the state of a classical system without prior knowledge of its state the result is also uncertain, but in a different way. If many systems $S_i$ are prepared in the same way and we measure the state of each one, all of our measurement devices $M_i$ will show the same result, just as in the case of the quantum preparation. Indeed, if the measurement devices are perfect, the only variation across devices stems from inaccuracies in the preparation of the systems $S_i$. But quantum uncertainty is different: even if Alice, say, diligently prepared all systems $S_i$ in the same exact way, Bob's measurement results could be highly uncertain, in the sense that he will observe a probabilistic distribution of results when the preparation had no such uncertainty. Where does this quantum uncertainty come from? What process could possibly create true randomness from certainty? 

\begin{figure}[htbp] 
   \centering
   \includegraphics[width=\columnwidth]{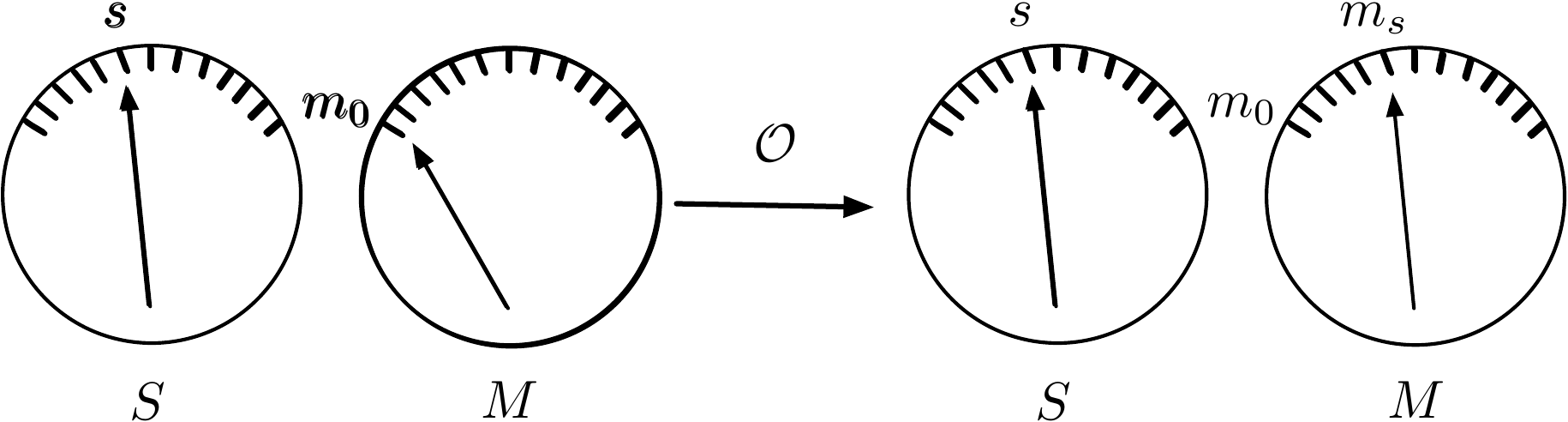} 
   \caption{In a classical measurement, the state of the system $S$ is ``copied" onto the measurement device so that the latter reflects the former. }
   \label{meas}
\end{figure}

Let's first take a look at how a classical measurement is supposed to work. Suppose you have a system $S$ that can take on different states $s$ (for ease of discussion we'll assume that the states are discrete), and a device $M$ that can take on states $m$. In order for the measurement device's state to reflect the system's state, the two need to become correlated. That is exactly what a good measurement operation will do, and that operation is essentially a ``copy" operation: If the measurement device is prepared in the state $m_0$ (some known calibrated state), then the measurement operation ${\cal O}$ should bring
\be
s\times m_0\stackrel{\cal O} {\rightarrow}s \times m_s\;.  \label{cmeas}
\ee
Here, $m_s$ is the ``copy" of $s$. It doesn't have to ``look" like $s$, but ideally there exists a one-to-one relationship between them (see Fig.~\ref{meas}).

\vskip 0.25cm
\noindent {\bf Quantum measurement}. The quantum measurement process is modeled after the classical idea, of course, but there is a fundamental difference: quantum states cannot be copied perfectly~\cite{Dieks1982,WoottersZurek1982}. Because of the linearity of quantum mechanics, the best we can do is to construct ``optimal cloning machines"~\cite{GisinMassar1997} that copy quantum states as accurately as is allowed by the laws of physics. Still, when measuring arbitrary states the measurement device will always be uncertain~\footnote{This minimum uncertainty so as not to violate the no-cloning theorem is, incidentally, the reason for the existence of Hawking radiation in black holes~\cite{AdamiVerSteeg2015}.}. Quantum mechanics provides us with a set of rules that allows us to predict this quantum uncertainty very accurately, and I will briefly outline those rules here because it is essential that we understand the respective roles of the system that is being measured $S$, and the measurement device $M$. In most accounts of measurement theory there is an asymmetry between $S$ and $M$ that will turn out to be artificial, but let's first make sure we distinguish the two. If we prepare a quantum system $S$  (I will focus on qubits here for simplicity, but all arguments can be extended to arbitrary dimension) at an angle $\theta$ with respect to the measurement device~\footnote{To keep things simple, we will ignore complex phases here.} $M$
\be
\vert\Psi \ra_S=\cos(\tht)\vert0\ra_S+\sin(\tht)\vert1\ra_S
\ee
and let it interact with a measurement ancilla (prepared in the state $\vert0\ra_M$) by way of the unitary measurement operator ${\cal U}$, the joint system becomes
\be
{\cal U}\vert\Psi\ra_S\vert0\ra_M=\cos(\tht)\vert0\ra_S\vert0\ra_M+\sin(\tht)\vert1\ra_S\vert1\ra_M\;. \label{entangle}
\ee
To quantify the statistics of such measurements, all we need to do is study the density matrix of the measurement device. 
In the present discussion we'll follow the system's wave function because
the time evolution of a quantum system is deterministic: it evolves in a unitary manner so that the joint probability distribution of a closed system is conserved. Probabilities emerge when we focus on subsystems of an entangled whole. For example, the density matrix of the measurement device $\rho_M$ is a mixed state even though the joint system of quantum state (including its preparation) and measurement device is pure. The density matrix of the measurement device is obtained by tracing the joint density matrix over the quantum system's Hilbert space~\footnote{A full description of the unitary theory of quantum measurement can be found in~\cite{GlickAdami2020}.}:
\be
\rho_M&=&
\cos^2(\tht)\vert0\ra_{\!M}\!\la 0\vert+\sin^2(\tht)\vert1\ra_{\!M}\!\la 1\vert \nonumber \\
 &=&\begin{pmatrix}\cos^2(\tht) & 0 \\0 & \sin^2(\tht)\end{pmatrix}\;. 
\label{rhom}
\ee
Equation (\ref{rhom}) tells us that Bob will be convinced he observed state `0' with probability $\cos^2(\tht)$ while he will record a `1' with probability $\sin^2(\tht)$. The states 0 and 1 are (it goes without saying) place holders for physical qubits, which could be spin-1/2 particles or the polarization states of a photon. For example, if the angle $\theta$ represents the rotation that a photon's polarization undergoes before it is detected by Bob, then the correlation between Alice's state preparation and Bob's detector is given by $\la M_AM_B\ra=\cos(\theta)$: if $\theta=90^{\degree}$, for example, Alice's measurement results (we can safely assume that Alice prepared her quantum states using measurements) and Bob's will be completely uncorrelated. 

The measurement device's density matrix (\ref{rhom}) is diagonal in its own basis (the ``pointer basis"), which means it can be considered a classical device.  The joint density matrix of the quantum system and the detector is also diagonal in this basis, but only because we have traced out Alice's measurement device that she used to prepare her states. In general, the joint density matrix of multiple measurement devices has off-diagonal terms, that is, it is non-classical~\cite{GlickAdami2020}. 

\begin{figure}[htbp] 
   \centering
   \includegraphics[width=0.3\columnwidth]{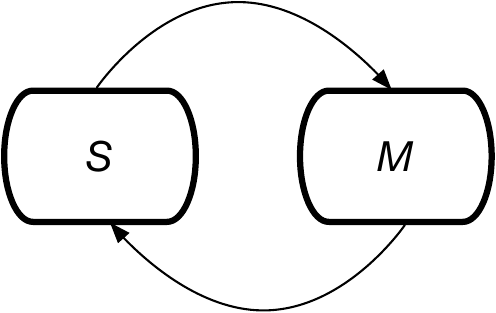} 
   \caption{In quantum measurement, system and device are entangled and measure each other.}
   \label{qm}
\end{figure}

Why is perfect quantum cloning impossible? At the very heart of this impossibility lies the fact that the entanglement operation (\ref{entangle}) is very different from the copy operation: it is linear, and it leads to a superposition of states. As a consequence, the distinction between system and device--what is measured and what is measuring--disappears. Fundamentally, system and device are measuring each other: they are one indivisible whole (see Fig.~\ref{qm}). 
Mathematically, the joint system after entanglement ``lives" in an entirely different Hilbert space, one that is twice as large as either the state's or the device's system. After the interaction, it takes twice as many bits to specify the final state, and there is just not enough ``room" in the measurement device to describe that joint space: the device simply doesn't have enough bits. I will argue here that {\em any} system (classical or quantum) that is forced to respond to a question without having enough bits to encode the answer will respond with some bits of randomness that are a consequence of logical undecidability. To make this connection, let me introduce undecidability more formally.

\vskip 0.25cm
\noindent{\bf Undecidability.}
In 1931, the Austrian mathematician Kurt G\"odel destroyed Hilbert's dream of a ``mechanical theorem solver"---the idea that it would one day be possible to construct an algorithm that could automatically prove all correct theorems. By formulating his eponymous ``Incompleteness Theorem", G\"odel showed that there cannot be a consistent set of axioms that can be used to prove all correct statements about natural numbers, implying that there are true statements about these numbers that will remain unprovable within the axiomatic system. Shortly after, Turing applied this sort of thinking to computer science~\cite{Turing1936} by showing that a computer program that can determine whether any particular computer program will ``halt" (meaning, terminate and issue a result) cannot exist: the halting probability is undecidable. 

One way to prove this result is to assume that such a program exists, and follow this assumption to a logical contradiction. It is important that we understand this contradiction for the arguments that follow. Suppose there is a function $F(s,m)$ that has two arguments: an input sequence $s$ and a program $m$. The function $F$ is supposed to return 1 if $m$ halts on $s$, and zero otherwise: it is a halting algorithm. Now imagine a program $M$ that takes a program $m$ and uses that program as its ``data": $M(m,m)$. Let's also say that $M(m,m)$ does the {\em opposite} of $F(m,m)$, that is, it returns 0 if $m$ halts on $s$ and 1 otherwise. Then by using a version of Cantor's algorithm it is now clear that the program $M$ cannot be among the set of programs $m$ (even though they are all enumerated), therefore the function $F(s,m)$ cannot exist. 

The reader may have noted my surreptitious use of $s$ and $m$ in the definitions of input and program, mirroring the notations for ``system" and ``measurement device" used earlier. Indeed it appears as if the impossibility of a halting algorithm is due to the fact that a program that evaluates itself {\em could} be made to be inconsistent with the state of the system as a whole. This is reminiscent of our earlier observation that a quantum measurement device cannot encode the joint state created after entanglement because the result of the measurement must be consistent with the system state $s$ which, after all, is also encoded in the joint state. In a way, we can say that a quantum system is asked to (in part) measure {\em itself}, because after measurement the only state that exists is the joint state of system and measurement device.

Is this similarity between the undecidability of the halting probability and quantum measurement just an accident? It certainly could be as the halting problem appears to pertain to Turing machines with infinite tapes, while quantum systems can be as small as single qubits. 
To study this question, we must first move away from ``the set of all programs of arbitrary length" that we used for Cantor's argument, and look at finite programs of length $n$. To do this, let's first review Chaitin's function $\Omega$. Say $S$ is the set of all programs and $H$ is the set of all programs that halt. Then Chaitin defines
\be
\Omega=\sum_{s\in H}2^{-\vert s\vert}\;,
\ee where $\vert s\vert$ denotes the size of $s$ (in bits). By Kraft's inequality $\Omega$ is a probability~\footnote{We must specify here that the programs are {\em self-delimiting}, which means that if a short program halts, we do {\em not} include in $\Omega$ all the programs that have that program as the beginning (so-called ``prefix-free'' codes).}, but because $F(s,m)$ does not exist (as we proved earlier), $\Omega$ is uncomputable. In fact, it is possible to prove that every bit of the binary representation of $\Omega$ is random, that is, $\Omega$ is not compressible~\cite{Chaitin1998,Chaitin1999}.  This randomness of $\Omega$ is a consequence of mathematical logic only. 

For finite programs of length $n$ the function $F(s,m)$ definitely exists (you cannot use Cantor's argument on finite sequences), and therefore we can calculate the halting probability $\Omega_n$ of all programs up to length $n$ (but not longer).
Let $S_n$ be the set of all programs of length up to $n$ and let $H_n$ be the set of all programs up to size $n$ that halt after running them for some time $T$ (the argument does not have to specify exactly how large $T$ is, but imagine that $T$ should scale with $n$). We can then define
\be
\Omega_n=\sum_{m \in H_n}2^{-\vert m\vert}\;.
\ee
If we wait long enough, all $n$ bits of $\Omega_n$ will be correct---meaning that they are the same bits as the first $n$ of $\Omega$'s bits~\footnote{Note that $\Omega_n<\Omega<\Omega_n +2^{-n}$ because (due to the prefix-free nature of the programs) all the halting programs longer than $n$ in the world can never add more than $2^{-n}$ to $\Omega_n$.}. In that case, the program that tests programs must have found {\em all} programs up to length $n$ that halt. This, in turn, implies that the program that encodes the algorithm $F(s,m)$ that achieves this feat {\em must} be longer than $n$, since otherwise a universal halting program would exist. This is significant because we have now shown that a program that attempts to determine if a copy of itself halts must be longer than the program itself: precisely the problem that we encountered in quantum measurement! 

What if we {\em force} a program of length $n$ to determine its own halting probability (as we do in quantum mechanics, as a matter of fact)? How will it respond? To answer this question, I will first construct simple machines that perform computations on binary sequences using binary programs, and then explore what happens if these machines are forced to operate on each other. This is not the same thing as asking a Turing machine to evaluate whether its own program will halt, but it implements the same dilemma: the outcome of a computation determines what computation will be performed, creating a possible paradox via self-reference.

\vskip 0.25cm
\noindent{\bf Entangled Classical Turing Machines.}
Let us construct machines that act on sequences of length $n$ and return a single bit as output (the construction can readily be generalized to write any number of output bits). This single output bit could be the $i$th bit of the halting probability $\Omega_n$, but for our purposes it could be {\em any} calculation returning a single bit. Such a computation is determined by one of $2^n$ logic tables (the computation is effectively an $n\to1$ logic gate), which we will encode as binary sequences $\vec m$, living in a space  $M_{2^n}$. This computation can be written formally using the linear operator ${\cal C}$
\be
{\cal C}=\sum_{m_1m_2\cdots m_{2^n}} P_{m_1m_2\cdots m_{2^n}}\otimes \sum_{i=1}^{2^n}P_i\otimes \sigma_x^{m_i}\;.\label{comput}
\ee 
Here, the operators $P_{m_1m_2\cdots m_{2^n}}$ are projectors acting on the space of programs (the vectors 
$\vert m_1\cdots m_{2^n}\ra$), while the $P_i$ are projectors on the input states $\vert s_1\cdots s_n\rangle$ (inputs to the logic operation specified by the program). The operator $\sigma_x^{m_i}$ acts on the output bit, conditionally flipping that bit (via the Pauli matrix $\sigma_x$) depending on whether $m_i=1$ (flip) or $m_i=0$ (don't flip). Because the state of the machine is fully determined by the joint program-input state, we will use the state of the bits as a proxy for the state of the Turing machines.

An explicit example of a pair of Turing machines implementing any of the four $1\to1$-computations (specified by a two-bit program) is given in the Appendix. 
In Eq.~(\ref{comput}) I've conveniently written the vectors $\vec s$ and $\vec m$ using the Dirac bra-ket notation, but because these represent classical states, they can never occur in superpositions in this context. 

A simple $2\to 1$ example computation is the CNOT (controlled NOT) operator ${\cal C}_{\rm CNOT}$ specified by program $\vert m\rangle=\vert0110\rangle$ acting on the set spanned by the vectors $\vert00\ra, \vert01\ra, \vert10\ra$ and $\vert11\ra$. So, for example to calculate CNOT(1,0), we apply ${\cal C}_{\rm CNOT}$ on the input state $\vert10\ra_s$, and pick the $m_i$ from the CNOT program $\vert0110\ra$. The term with projector $P_{10}$ is the only non-vanishing contribution to ${\cal C}_{\rm CNOT}$, and looking up the third bit in the program $\vert0110\ra$ tells us that the ancilla needs to be flipped for this particular input (we suppress the program vector here for simplicity):
\be
{\cal C_{\rm CNOT}}\vert10\ra_s\vert0\ra=\vert10\ra_s (\sigma_x)^1\vert0\ra=\vert10\ra_s \vert1\ra  \label{cnot}\;.
\ee
A particular important operation (program) is the simplest logic operation acting on a single bit: the operation COPY. We can write this as
\be
{\cal O}= P_0\otimes \one+P_1\otimes \sigma_x\;, \label{copy}
\ee
with $P_0=\vert0\ra\la0\vert$ and $P_1=\vert1\ra\la1\vert$. When writing (\ref{copy}), I have suppressed the ``copy"  program that ${\cal O}$ is conditional on, namely the program $\vert01\ra$.  This operation is precisely the one that implements the measurement process (\ref{cmeas}) on classical bits, so that
\be
{\cal O}\vert0\ra\vert0\ra&=&\vert0\ra\vert0\ra\\
{\cal O}\vert1\ra\vert0\ra&=&\vert1\ra\vert1\ra\;.
\ee
In fact, this turns out to be the classical version of the quantum operator ${\cal U}$ in Eq.~(\ref{entangle}): they are given by precisely the same equation, and only differ in the states that the operation is applied to (the classical operator is never applied to superpositions). 

We now start considering the possibility that the bits that the Turing machine conditions on (input bits and program bits) might be modified by another Turing machine. In this manner, both Turing machines are entangled with each other, but classically of course. For this to be possible we must account for the possibility that the measurement device is {\em not} in the prepared state $\vert0\ra$. This is not a problem in principle, because we can simply extend the range of the operator ${\cal O}$ so that
\be
{\cal O}\vert0\ra\vert1\ra&=&\vert0\ra\vert1\ra\\
{\cal O}\vert1\ra\vert1\ra&=&\vert1\ra\vert0\ra\;.
\ee
Measurements with an uncertain initial state are not uncommon in quantum mechanics, where the degree of uncertainty of the ancilla simply implements different strengths of measurements~\cite{Aharonovetal1988,Brun2002}. Allowing for unprepared ancilla states begins to blur the distinction between system and measurement device. 

\begin{figure}[htbp] 
   \centering
     \includegraphics[width=\columnwidth]{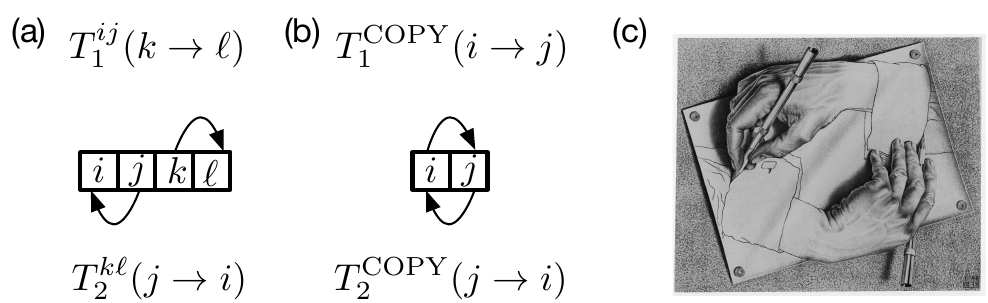} 
   \caption{(a) Two machines $T_1$ and $T_2$ that write on each other's program spaces. $T_1$ uses bits $i$ and $j$ as programs, to operate on bits $k$ and $\ell$, while $T_2$ uses the program in bits $k$ and $\ell$ with I/O bits $j$ and $i$ (this machine is constructed in detail in the Appendix). Other permutations of the bits $i,j,k,\ell$ are equally possible. (b) Two machines with fixed COPY programs, writing on each other's I/O bits. (c): Two non-universal entangled machines (with permission from the M.C. Escher Company.) } 
   \label{turing}
\end{figure}

The simplest general scenario of two Turing machines $T_1$ and $T_2$ implementing $1\to1$ logic and operating on each other's program space would involve four bits: two of which serve as program space, as well as one for input and one for output). The I/O space of one of the Turing machines would serve as the program for the other, and vice versa. In such a scenario, it is immediately clear that the state of each of the systems $s$ or $m$ is not well-defined, because which operation $T_1$ will perform on its I/O space will depend on what operation $T_2$ is performing, since $T_1$'s program is determined by the operation of $T_2$. What that operation is, however, depends on $T_1$ (see Fig.~\ref{turing}a).
Representing such a pair of transformations mathematically using the formalism from Eq.~(\ref{comput}) is possible but cumbersome (this is carried out in the Appendix).  To obtain some insights on the consequences of potentially incompatible operations on classical bits, I will simplify the situation even further: Let's keep the program of the two Turing machines unchanged (the simple COPY operation (\ref{copy})), but the bit that $T_1$ writes to is the bit that simultaneously $T_2$ reads from, and vice versa (Fig.~\ref{turing}b). 
 
The copy operators $U$ and $V$ copying the ``left" bit to the ``right" bit (and vice versa) can be written as
\be
U&=&P_0 \otimes \one +P_1\otimes \sigma_x \\
V&=& \one \otimes P_0+ \sigma_x \otimes P_1\;.
\ee
Using those, we find that
\be
UV\vert00\ra=\vert00\ra  &\ \ \  &    VU\vert00\ra = \vert00\ra\\
UV\vert01\ra=\vert10\ra & & VU\vert01\ra=\vert11\ra\\
UV\vert10\ra=\vert11\ra & & VU\vert10\ra=\vert01\ra\\
UV\vert11\ra=\vert01\ra & & VU\vert11\ra=\vert10\ra\;.
\ee
It is clear that the order of operations matters, except for the input state $\vert00\ra$. For inputs $\vert01\ra$ and $\vert10\ra$ one of the bits is inconsistent (between the two orders), while for the input state $\vert11\ra$ both bits are inconsistent. But it is also clear that when acting on states with operators, we can really only implement one or the other operation at a time (thus performing operations sequentially), and doing so does not create ambiguities.  In fact, even simulating {\em simultaneous} operations on a computer is impossible because the computer must also operate sequentially. Generally speaking, simultaneity is incompatible with classical physics, and not just because the order of events can change depending on the frame of reference (relativity). Only in quantum physics is the act of $S$ measuring $M$ identical with the act of $M$ measuring $S$, because both of these measurements create the same entangled state. But what kind of state is created by acting with $UV$ or $VU$ on an arbitrary input state?

One way to answer this question is to move away from a deterministic sequential picture by writing the transformations statistically, in terms of their effect on a (classical) density matrix instead that represents a uniform probabilistic mixture of the four possible input states.
If we choose the representation $\vert0\ra=\begin{pmatrix} 0\\1\end{pmatrix}$ and $\vert1\ra=\begin{pmatrix} 1\\0\end{pmatrix}$, we can write this mixed state as the matrix
\be
\rho_{\rm in}=\vert{\rm in}\ra\la {\rm in}\vert=\frac14\begin{pmatrix}1 &0 &0&0\\ 0 &1 &0&0\\0 &0 &1&0\\0 &0&0&1\end{pmatrix}\;.
\ee
The transformations $U$ and $V$ in this representation read
\be
U=\begin{pmatrix}0 &0 &1&0\\ 0 &1 &0&0\\1 &0 &0&0\\0 &0 &0&1\end{pmatrix}\; \;
V=\begin{pmatrix} 0 &1 &0&0\\1 &0 &0&0\\0 &0 &1&0\\0 &0 &0&1\end{pmatrix}\;.
\ee
The matrices $U$ and $V$ are symmetric orthogonal matrices (so that $UU^T=VV^T=\one$) but with negative determinant: they are reflections. Both the matrices $UV$ and $VU$ are non-diagonal: in fact they are elements of the special orthogonal group in four dimensions, SO(4)\footnote{This is not particularly surprising, as the quantum counterpart of these operations (the measurement/entanglement operator ${\cal U}$ for qubits shown in (\ref{entangle})) belongs to the group SU(2)$\times$SU(2), which is a double cover of SO(4).}.  Because of the symmetry of $U$ and $V$,  $(UV)^T=VU$. Furthermore, acting with $UV$ twice is the same as acting once with $VU$, that is, $(UV)^2=VU$. We call $UV$ the ``left" operation, while $VU$ is the ``right" operation, which are only different here because of classical physics. 

We can get some insights into what kind of state is created by acting with $UV$ or $VU$ on an arbitrary input state by looking at the spectrum of these operators (they have the same spectrum). Besides the two eigenvalues $\lambda_{1,2}=1$, they also have a pair of complex eigenvalues $\lambda_{3,4}=-\frac12(1\pm i\sqrt{3}$), which seems to imply that the final state of the two bits these operators act on are undetermined.

Let us now attempt to transform $\rho_{\rm in}$ with $UV$ or $VU$. Operating with $U$ or $V$ does not change the input state because $U\rho_{\rm in}U^T=
V\rho_{\rm in}V^T=\rho_{\rm in}$. The same holds of course for $UV\rho_{\rm in}(UV)^T$.  What if we operate with $UV$ from the left, and with $VU$ from the right?
\be\rho_{\rm in}\stackrel{?}{\longrightarrow}\rho_{\rm out}^{(L)}=UV\rho_{\rm in}(VU)^T   \label{ill}\ee
This is, to put it mildly, an illegal operation: it does not result in a density matrix because the operation is not trace-preserving. Nevertheless, it does appear to create both transformations at the same time as we shall see. 

Let us investigate the ramifications of this transformation by considering the ``left" and ``right" output states
\be
\rho_{\rm out}^{(L)}=UV\rho_{\rm in}(VU)^T\;\;\;\rho_{\rm out}^{(R)}=VU\rho_{\rm in}(UV)^T\;.
\ee
Because $\rho_{\rm in}$ is diagonal, the matrices $UV$ and $VU$ are really just performing a singular value decomposition (SVD) of the transformed (output) state, so (keeping in mind that $(UV)^2=VU$ and $(VU)^2=UV$)
\be
\rho_{\rm out}^{(L)}=\frac14VU;\;\; \rho_{\rm out}^{(R)}=\frac14UV\;.
\ee
As the spectrum of these matrices is just 1/4 of the spectrum of $UV$ and $VU$, the same statement about the undecidability of the two output bits applies. However, while $\tr\rho_{\rm in}=1$,  $\tr\rho_{\rm out}^{(L)}=\tr\rho_{\rm out}^{(R)}=\frac14$, that is, it appears we have ``lost" probabilities. The loss of probability is not a problem in principle as it is a sign of dissipative dynamics; but in a closed system this is inadmissible in principle, and can be traced back to the complex eigenvalues that signal the undecidable, never-halting, state.

But let us cast such concerns aside for now and forge ahead for a little while longer. What is the density matrix for each of the two bits separately? Surely if we ask about the state of a single bit we will receive an answer? We can trace over the ``other" bit, respectively, to see what that answer might be. For the ``left" matrix we find
\be
\rho_1&=&\tr_2(\rho_{\rm out}^{(L)})= \frac14\begin{pmatrix} 0 & 0\\ 1 & 1\end{pmatrix}=\frac14\vert0\ra\Bigl(\la0\vert+\la 1\vert\Bigr)
\;,\\
\rho_2&=&\tr_1(\rho_{\rm out}^{(L)})=\frac14\begin{pmatrix} 0 & 1\\ 0 & 1\end{pmatrix}=\frac14\Bigl(\vert0\ra +\vert1\ra\Bigr)\la 0\vert\;.
\ee

Obviously, both matrices are non-diagonal. They each have two eigenvalues 0 and 1/4, and in a sense they ``mirror each other", just as the bits in Fig.~\ref{turing}(b) do. For $\rho_1$ (describing the leftmost of the two bits) the right eigenvector to the eigenvalue 1/4 is the state $\vert0\ra$. The right eigenvector of $\rho_1$ with vanishing eigenvalue is, on the other hand, the state $\vert0\ra-\vert1\ra$: a superposition!
The mirror situation holds for the ``right" of the two bits. The {\em left}  eigenvector with eigenvalue 1/4 is the state $\vert0\ra$, while the left eigenvector (with vanishing eigenvalue) is again the superposition $\vert0\ra-\vert1\ra$. If we had used the $UV$ matrix to perform the ``simultaneous" transformation (giving rise to the output $\rho_{\rm out}^{(R)}=\frac14UV$), the only effect would the that $\rho_1$ and $\rho_2$ switch labels (i.e., the left bit becomes the right bit, and vice versa). They are, after all, transposes of each other. 

Perhaps the reader is at this point shaking their head vigorously, as I promised that superpositions would not appear in the classical description. And to some extent they do not: these eigenvectors in superposition are just ghosts, as they appear with vanishing strength (zero probability). They are, if you will, mere palimpsests of the quantum case, announcing their possible presence only by insinuation.
Indeed, it is not the presence of these superpositions that is an abomination, what is {\em really} forbidden is the classicality of these bits. After all, classical physics does not exist: we just find it easier to conceptualize our reality this way. 

The fact that the superposition of states occurs with zero probability essentially tells us that the computations that we asked these Turing machines to perform will never halt, but loop endlessly over the possibilities. In real physics (quantum physics, that is) these computations do halt, but the value that these bits take on--like the bits of the halting probability--must be random because if they were not then halting algorithms shorter than program size could be written. In hindsight, these random bits that are forced upon us by the ``strange loop" of quantum measurement (measurement devices measuring each other, where the distinction between system and device has fully lost its meaning) represent the {\em only} reality. What we perceive to be real in our world are just conditional probabilities relating such random numbers to each other: the only reality that is accessible to us is the relative state of measurement devices~\cite{Rovelli2015}. So, while to Einstein it appeared that quantum mechanics must be incomplete~\cite{Einsteinetal1935}, it now seems as if he glimpsed the incompleteness of mathematics instead.

\subsection*{Acknowledgements} I am grateful to numerous people that have patiently listened to my ruminations about the link between quantum and classical entanglement, their relation to the halting problem, and the origin of quantum indeterminacy. First and foremost is Arend Hintze, who hammered out the Turing machine representation of logic operations with me, many years ago, and attempted to simulate entangled classical Turing machines on a (classical) computer (to no avail). I've also discussed these concepts at length with Richard Lenski. Thanks are also due to Thomas Sgouros, whose comments on the manuscript helped improve it.  Finally, I am grateful to Nicolas Cerf for a decade of collaboration on quantum information theory, which laid the foundations for these ideas, and in particular to Greg Chaitin, without whose phone call in the late 1990s to my office at Caltech none of this work would exist. An earlier version of this work was submitted to the FQXi essay contest. 

\section*{Appendix}
\begin{figure}[t] 
   \centering
   \includegraphics[width=0.75in]{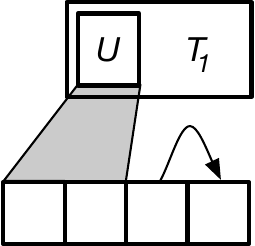} 
   \caption{A machine $T_1$ reading a two-bit program $U$ (from the two left-most bits) that controls a $1\to1$ read-write operation on the two right-most bits.}
   \label{fig:T1}
\end{figure}

In the main text I discussed a construction where two Turing machines~\footnote{These machines are not universal in the sense of Turing, nor is this required for these arguments.} with fixed programs write on each other's I/O space, leading to inconsistent bits without a determined state. But because the programs were fixed for simplicity, this leaves the possibility that the programs' indeterminacy could compensate for the I/O indeterminacy. Indeed, for this reason, the joint state of program and data in the main text was itself indeterminate. Here I discuss the case where the programs of the Turing machines are not fixed, but can be modified via writing into that space. In particular, I show that in the minimum system of programs of size 2 and an I/O space of size two (see Fig.~\ref{fig:T1}), classical indeterminacy occurs on the level of the individual machines, while the joint state (program and I/O space) is determined.

\begin{figure}[htbp] 
   \centering
   \includegraphics[width=0.75in]{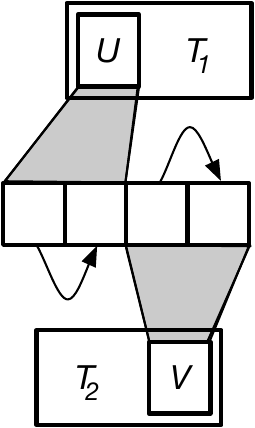} 
   \caption{The machine $T_2$ reads its program $V$ from those bits that are written by program $U$ of machine $T_1$, which in turn depends on the program $U$. As a consequence, the state of some or all of  the bits could be undecidable.}
   \label{fig:T12}
\end{figure}

There are four possible programs for machine $T_1$ acting on the 2-bit I/O space, given by
\be
U_{00}&=& P_0\otimes \one +P_1\otimes \one \label{four-1} \\
U_{01}&=& P_0\otimes \one +P_1\otimes \sigma_x\\
U_{10}&=& P_0\otimes \sigma_x +P_1\otimes \one\\
U_{11}&=& P_0\otimes  \sigma_x +P_1\otimes  \sigma_x\;, \label{four-4}
\ee
specified by the four projectors $P_{00},P_{01},P_{10},P_{11}$ acting on the 2-bit program space ($P_{ij}=\vert ij\ra\la ij\vert$), that is,
\be
T_{1}=\sum_{m_1m_2}P_{m_1m_2}\otimes U_{m_1m_2}\;.
\ee
Figure \ref{fig:T1} shows a machine implementing one of the four operations in Eqs.~(\ref{four-1}-\ref{four-4}) (depending on the bits on the tape) on the I/O bits.  Now we imagine a machine $T_2$ that reads its program from the two right-most bits instead, controlling the $1\to1$ read-write logic in the two left-most bits (left-to-right). These programs $V_{ij}$ are the same as the programs $U_{ij}$ (but the operator acts on bits 1 and 2 instead) 
and implements the machine (see Fig.~\ref{fig:T12})
\be
T_{2}=\sum_{m_1m_2} V_{m_1m_2}\otimes P_{m_1m_2}\;.
\ee
Choosing the representation $\vert0\ra=\begin{pmatrix} 0\\1\end{pmatrix}$ and $\vert1\ra=\begin{pmatrix} 1\\0\end{pmatrix}$ as in the main text, the operators $T_1$ and $T_2$ are $16\times 16$ matrices $U$ and $V$ (acting on the same four-bit vector space) with positive unit determinant. They are members of the group ${\rm SO}(16)$ (specifically, they are permutations). Acting on the density matrix of a random 4-bit state $\rho_{\rm in}=\frac1{16}\one$ we can readily derive the output states ($\rho_{\rm out}^{(L)}=(\rho_{\rm out}^{(R)})^T$)
\begin{widetext}
\be
\rho_{\rm out}^{(R)}=VU\rho_{\rm in}( UV)^T=\frac1{16}\left(
\begin{array}{cccccccccccccccc}
 0 & 0 & 0 & 0 & 1 & 0 & 0 & 0 & 0 & 0 & 0 & 0 & 0 & 0 & 0 & 0 \\ 
 0 & 1 & 0 & 0 & 0 & 0 & 0 & 0 & 0 & 0 & 0 & 0 & 0 & 0 & 0 & 0 \\ 
 0 & 0 & 0 & 0 & 0 & 0 & 1 & 0 & 0 & 0 & 0 & 0 & 0 & 0 & 0 & 0 \\ 
 0 & 0 & 0 & 1 & 0 & 0 & 0 & 0 & 0 & 0 & 0 & 0 & 0 & 0 & 0 & 0 \\ 
 1 & 0 & 0 & 0 & 0 & 0 & 0 & 0 & 0 & 0 & 0 & 0 & 0 & 0 & 0 & 0 \\ 
 0 & 0 & 0 & 0 & 0 & 1 & 0 & 0 & 0 & 0 & 0 & 0 & 0 & 0 & 0 & 0 \\ 
 0 & 0 & 1 & 0 & 0 & 0 & 0 & 0 & 0 & 0 & 0 & 0 & 0 & 0 & 0 & 0 \\ 
 0 & 0 & 0 & 0 & 0 & 0 & 0 & 1 & 0 & 0 & 0 & 0 & 0 & 0 & 0 & 0 \\ 
 0 & 0 & 0 & 0 & 0 & 0 & 0 & 0 & 0 & 0 & 0 & 0 & 1 & 0 & 0 & 0 \\
 0 & 0 & 0 & 0 & 0 & 0 & 0 & 0 & 0 & 0 & 0 & 0 & 0 & 1 & 0 & 0 \\ 
 0 & 0 & 0 & 0 & 0 & 0 & 0 & 0 & 0 & 0 & 1 & 0 & 0 & 0 & 0 & 0 \\ 
 0 & 0 & 0 & 0 & 0 & 0 & 0 & 0 & 0 & 0 & 0 & 1 & 0 & 0 & 0 & 0 \\ 
 0 & 0 & 0 & 0 & 0 & 0 & 0 & 0 & 1 & 0 & 0 & 0 & 0 & 0 & 0 & 0 \\ 
 0 & 0 & 0 & 0 & 0 & 0 & 0 & 0 & 0 & 1 & 0 & 0 & 0 & 0 & 0 & 0 \\ 
 0 & 0 & 0 & 0 & 0 & 0 & 0 & 0 & 0 & 0 & 0 & 0 & 0 & 0 & 1 & 0 \\
 0 & 0 & 0 & 0 & 0 & 0 & 0 & 0 & 0 & 0 & 0 & 0 & 0 & 0 & 0 & 1 \\ 
\end{array}
\right)\;.\ee
\end{widetext}
While this matrix is non-diagonal, it only has real eigenvalues. We therefore can say that the (joint) state of the two machines is determined.

The density matrix of the left-most ($\rho_{1,2}$) or right-most ($\rho_{3,4}$) two bits can be obtained by tracing over the other two (they are the transpose of each other):
\be
\rho_{1,2}=\tr_{3,4} (\rho_{\rm out}^{(R)})=\frac1{16}\left(
\begin{array}{cccc}
0 & 0 & 0 & 0\\ 
1 & 1 & 0 & 0\\ 
0 & 0 & 2 & 0\\ 
0 & 0 & 2 & 2\\ 
\end{array}\right)\;.
\ee
As in the simpler two-bit case, this matrix is not unit trace (it is not a density matrix anymore), and $\rho_{1,2}$ has one (right) eigenvector that is a superposition: it is $\vert1\ra(\vert0\ra-\vert1\ra)$, also with eigenvalue zero. This is also the (left) eigenvector with zero eigenvalue for $\rho_{3,4}$. Thus, each of the individual machines is in an undetermined state while the joint state is determined: precisely mirroring the situation we encounter in quantum measurement.
%

%

\end{document}